\documentclass[twocolumn,showpacs,preprintnumbers,amsmath,amssymb]{revtex4}

\usepackage{graphicx}
\usepackage{dcolumn}
\usepackage{bm}

\begin{document}

\title{Non Poisson intermittent events in price formation}
\author{Antonella Greco$^1$, Luca Sorriso-Valvo$^1,2$, Vincenzo Carbone$^1$}
\address{$^1$Dipartimento di Fisica and Istituto Nazionale di Fisica della Materia, \\
Universit\`a della Calabria, Ponte P. Bucci Cubo 31C, 87036 Rende (CS),
Italy. \\
$^2$LICRYL - INFM/CNR, Universit\`a della Calabria, Ponte P. Bucci Cubo 31C, 87036 Rende (CS) - Italy.}

\begin{abstract}

The formation of price in a financial market is modelled as a chain of Ising spin with three fundamental figures of trading. We investigate the time behaviour of the model, and we compare the results with the real EURO/USD change rate. By using the test of local Poisson hypothesis, we show that this minimal model leads to clustering and ``declustering'' in the volatility signal, typical of the real market data. 

\end{abstract}


\maketitle

\begin{figure}
\includegraphics[width=9cm]{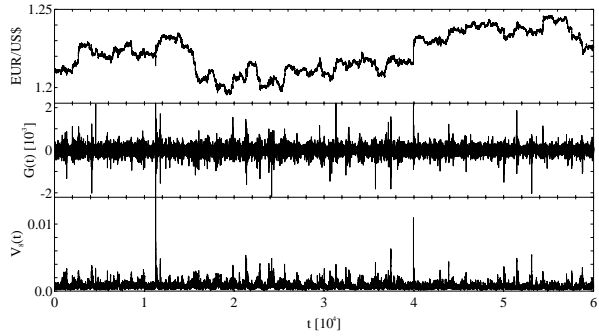}
\caption{We report the time evolution of the exchange rate EURO/USD $e(t)$ (upper panel), the return $G(t)$ with $\Delta t=1$min (middle panel) and the volatility $V_T(t)$ for $T=8$min (lower panel).}
\label{fig1}
\end{figure}

\begin{figure}
\includegraphics[width=8.6cm]{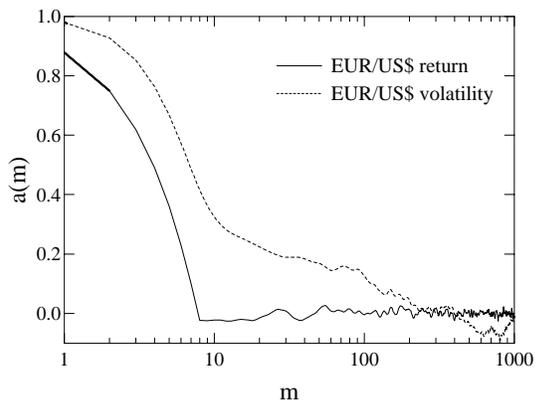}
\caption{The solid line is the autocorrelation function for returns of the exchange rate EURO/USD displayed in lin-log scale. The dashed one represents the autocorrelation function for the volatility of the same signal in the same scale.}
\label{fig2}
\end{figure}

\begin{figure}
\includegraphics[width=8.6cm]{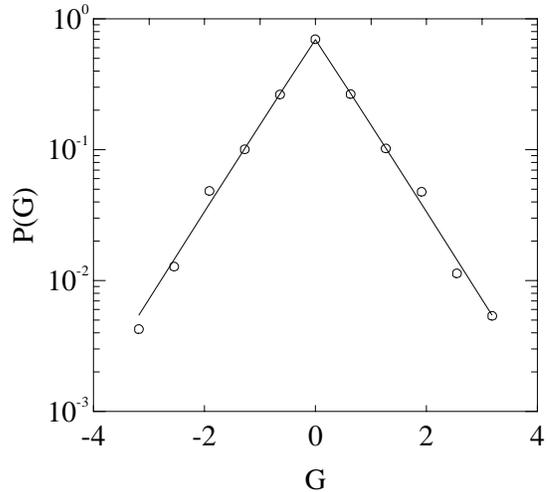}
\caption{Standardized PDF for returns $G(t)$ of the EURO/USD exchange rate. The solid line is the stretched exponential fit with with $c=1.0\pm0.1$.}
\label{fig3}
\end{figure}

\begin{figure}
\includegraphics[width=9cm]{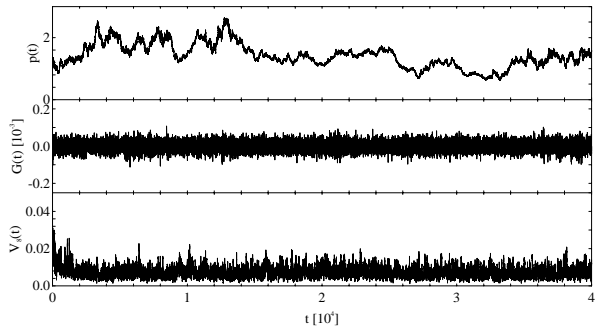}
\caption{We report the time evolution of the stock price $p(t)$ (upper panel), the return $G(t)$ with $\Delta t=1$ (middle panel) and the volatility $V_T(t)$ for $T=10$ (lower panel).}
\label{fig4}
\end{figure}

\begin{figure}
\includegraphics[width=8.6cm]{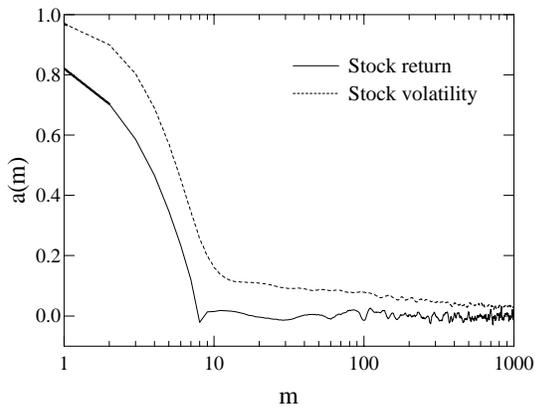}
\caption{The solid line is the autocorrelation function for returns of the stock price displayed in lin-log scale. The dashed one represents the autocorrelation function for volatility of the stock price in same scale.}
\label{fig5}
\end{figure}

\begin{figure}
\includegraphics[width=8.6cm]{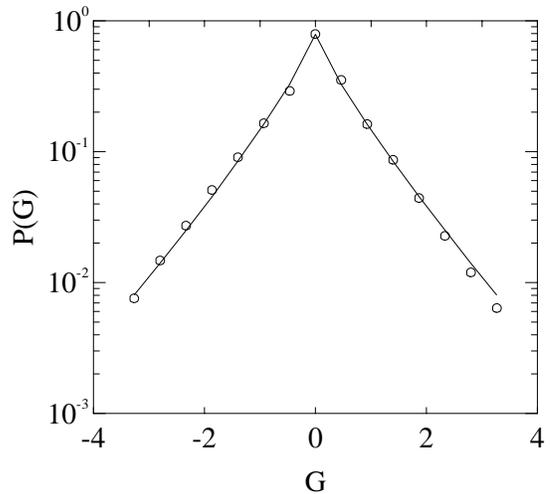}
\caption{Standardized PDF for returns $G(t)$ of the stock price. The solid line is the stretched exponential fit with with $c=0.85\pm 0.01$.}
\label{fig6}
\end{figure}

\begin{figure}
\includegraphics[width=8.6cm]{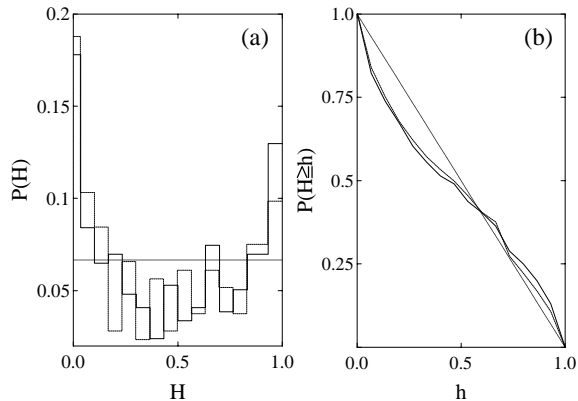}
\caption{In the panel (a) probability $P(H)$ of exchange rate EURO/USD volatility (solid line) and of stock price volatility (dashed line) as a function of $H$ is shown. A uniformely distribution $P(H)$ in $[0;1]$ is also shown as a reference (dotted line). In the panel (b) cumulative probability $P(H \geq h)$ of exchange rate EURO/USD volatility (solid line) and of stock price volatility (dashed line) as a function of $h$ is shown. $P(H \geq h)=1-h$ is also shown as a reference (dotted line).}
\label{p_h}
\end{figure}

Despite the obvious interest to adequately describe the stochastic behaviour of price formation in a financial market \cite{econofisica}, stochastic modelling misses some interesting, and perhaps crucial characteristics of the real market. In particular, it is well known that the market cannot be adequately described by time series of independent and identically distributed realizations of a random variable. In fact time series in the financial framework depend on a large number of strongly interacting systems. Then, the problem should be investigated in the framework of complex systems. In this case long-range correlations, not obvious \textit{a priori} and very difficult to be adequately modelled, are present in the system. In this situation every step towards the goal of analysing and describing correlations is welcomed. The Ising spin system \cite{ising} is one of the most frequently used models in statistical mechanics. Being made simply by binary variables, the model is able to reproduce different complex phenomena in different areas of science like economy \cite{econo0,econo1,econo2}, biology \cite{bio} and sociology \cite{socio}, were many interactions are required among various discrete elements. Some times ago a model of this kind has been used to describe the decision making mechanism of a closed community \cite{K2000}, where, in contrast to the usual majority rules \cite{majo}, the influence was spreading outwards from the center.\\
In this paper we use the Ising spin as a minimal model of financial market, in order to investigate and reproduce the non-Poisson characteristics of isolated events in price formation.
Besides the statndard statistical tools, we use a local analysis of time intervals between volatility events, which is not affected by the non-stationarity of the signal.\\
The dataset we investigate is the exchange rate EURO/USD collected every minute during two months in the period from May 25, 2004, up to July 31, 2004. The exchange rate $e(t)$ is represented by a discrete time series where $t = t_k = k \Delta t$ ($\Delta t$ is the sampling rate and $k = 0,1, \dots, N$). The main quantities of interest for traders within the market is the return of price $r(t)$, defined as the change of the logarithmic price over a time interval $\tau$, namely $G(t) = \ln e(t+\tau) - \ln e(t)$. A measure of the magnitude of fluctuations of prices used by traders is the volatility, defined as the absolute values of returns averaged over a window of time extension $T = n \Delta t$
\begin{equation}
V_T(t)=\frac{1}{n} \sum_{t^\prime =t}^{t+n-1} | r(t^\prime)|
\label{volatility}
\end{equation}
where $n$ is an integer representing the moving average window size. A further quantity is the autocorrelation function of returns (or of volatility) defined as 
\begin{equation}
a(r,m) = \frac{\sum_{n = m + 1}^{N} \left[G_n- \langle G \rangle \right]\left[G_{n-m}-\langle G \rangle \right]}{\sum_{n=1}^N \left[G_n-\langle G \rangle \right]^2}
\label{autocorr} 
\end{equation}
where $G_n = G(t_n)$. The main characteristics of the time serie are summarized in Figure \ref{fig1} where we show the time evolution of various quantities, namely $e(t)$, $G(t)$ and $V_T(t)$. We note the typical non-stationarity, the presence of bursty and explosive events in the return dataset and of isolated volatility events.\\
We also report the autocorrelation function for both the returns and the volatility (see Figure \ref{fig2}). While the former displays a very fast decay ($m_\tau \leq 10$ min), the latter displays a much longer persistent time lag.\\ 
Finally in Figure \ref{fig3} we report the standardized probability density function (PDF) of returns, which shows very clearly typical ``fat" tails.\\
The above characteristics can be adequately modelled by the Ising spin chain. The community of interacting agents which partecipate to the market, is modelled as a chain of one-dimensional Ising spins $s_i$ ($i = 1, \dots, N$). Each agent, at a time $t$, has an attitude to place buy orders, described by an up-spin $s_i(t) = 1$, or to place sell orders which is described by a down-spin $s_i(t) = -1$. The price of the asset at a given time $t$ is proportional to the normalized difference between demand and supply
\begin{equation}
y(t) = \frac{1}{N} \sum_{i=1}^N s_i(t)
\label{price}
\end{equation}
which in the classical Ising model is the magnetization. Here $|y(t)| \leq 1$ represents a probability. The evolution of the Ising chain is controlled by three simple rules that describe three figures partecipating to the market as in Ref.~\cite{K2000}:\\
1) in general a lot of market agents follows the trend. This is described by the rule that both agents in two next neighbouring sites $(i-1,i+2)$ follow the trend given by agents placed in $(i,i+1)$, namely when $s_is_{i+1} = 1$ then $s_{i-1} = s_{i+2} = s_i$.\\
2) Some agents prefer to act on the market independently on the neighbour agents. Then in absence of a precise ``tendency", sites $(i-1,i+2)$ are chosen at random, namely when $s_is_{i+1} = -1$ then both $s_{i-1}$ and $s_{i+2}$ are randomly chosen in an independent way.\\
3) A third figure is partecipating to the market, that is an agent with some knowledge external to the market. This is known in literature as ``fundamentalist" \cite{econo1,econo2}. This figure has an exact knowledge of what happens in the market so that he knows the difference between demand and supply. If supply is greater than demand he places a buy order, while he places a sell order when the demand is greater than supply.\\
Realizations of the Ising spin model have been analysed starting from a random sequence of $N = 1000$ spins at $t = 0$. Each interaction between two agents is interpreted as the tick-by-tick interaction between traders, while the price fixed at a given time $t$, defined as $p(t)= \exp \left[y(t)\right]$, is calculated after a single evolution of the whole chain.
Of course time is arbitrary in the Ising model. Actually, we have varied the value of the sampling rate in the numerical dataset and we have compared it with the real dataset. On the basis of the results obtained fron the simulation, we can suppose that each update of the whole chain represents the price every minute of the real market. Then, in the following, $p(t)$ is proportional to $e(t)$.\\
As in Ref.~\cite{K2000}, we compare the real data of EURO/USD exchange rate with the numerical results of the simulation.
The statistical analysis of $p(t)$ shows the same features of real exchange $e(t)$, namely the non-stationarity of the signal (see Figure \ref{fig4}), the behaviour of the correlation functions (see Figure \ref{fig5}) and, finally, the fat tailed PDFs of returns, as shown in Figure \ref{fig6}.\\ 
The statistical analysis carried on until now shows that this minimal model of financial market can capture 
many features similar, in a statistical sense, to the real market ones.\\
To further characterize the temporal properties of these time series (exchange rate and stock price) and investigate on the characateristics of isolated events in pricing formation, we analyze the distribution of waiting times (WT) between these events.\\
The statistical distribution of WT between discrete or isolated ``events" has been shown to be a powerful tool to investigate and characterize temporal point processes like solar flares \cite{B99}, earthquakes \cite{quakes}, dissipative events in turbulent flows \cite{squadra}, disruptions in laboratory plasmas \cite{plasma} and so on \cite{generico}. We consider the volatility of a financial market as a temporal point process with a certain number of discrete ``events", as the spikes in Figures \ref{fig1} and \ref{fig4}.\\
To quantify these events, we need to fix a threshold, above which we can select a spike as an ``event". We adopted a method according which all events that are above a certain threshold are cut. Then, we calculate the flatness of the remaining signal. The value of the threshold is varied, until the flatness of the remaining signal is equal to the Gaussian value $3$ \cite{O00}.
At this point we calculate the waiting times from the difference between the end time and start time of two next events (quiet times).\\ 
The statistical analisys of such variables can be used to detect the presence of long-range correlations in the dataset. For example, the study of the autocorrelation function of quiet times of long-term persistent records can provide informations about the presence or not of clustering in the extreme events ~\cite{generico,clustering}. Due to the non-stationarity and the poor statistics of our dataset ($\sim 300$ events) this kind of analisys can not be performed. However, a local Poisson hypothesis test can be used to check for presence of clustering in short and non-stationary records ~\cite{lepreti,nuovocimento}.\\
Let us consider the time sequence of extreme events selected in the volatility data according the rules explained above. Let us suppose that each event occurs at the discrete time $t_i$; each $\imath$th event can be associated to the two nearest and next nearest neighboring events, which occur at the times $t_{i-1}$, $t_{i+1}$, $t_{i-2}$ and $t_{i+2}$, respectively. Let $\delta t_i = \min \{t_{i+1}-t_i; t_i-t_{i-1}\}$ and let 
$\delta \tau_i$ be either $=t_{i+2}-t_{i+1}$ (if $\delta t_i =t_{i+1}-t_i$) or $\delta \tau_i=t_{i-1}-t_{i-2}$ (if $\delta t_i =t_i-t_{i-1}$). If a local Poisson hypothesis holds, both $\delta t_i $ and $\delta \tau_i$ are indipendently distributed according two Poissonian probability densities: $P(\delta t_i)=2\lambda_i\exp(-2\lambda_i \delta t_i)$ and $P(\delta \tau_i)=\lambda_i\exp(-\lambda_i \delta \tau_i)$. In this case, the stochastic variable $H$ defined as 

\[
H = {\delta t_i \over \delta t_i+\frac{1}{2}\delta \tau_i}
\]
is uniformely distributed in $[0;1]$ and it has a cumulative distribution $P(H\geq h) = 1-h$ independent on $\lambda$. In a process, where $\delta \tau_i$s are systematically smaller than $2\delta t_i$s, much more events are found respect to the Poissonian case, that is clusters are present and the average value of $H$ is greater than $1/2$. Conversely, a value of $\langle H \rangle$ less than $1/2$ means that voids are present in the signal.
The values of $H$ have been calculated from the volatility dataset of the change rate EURO/USD and from the stock price $p(t)$ obtained from the Ising model. In Figure \ref{p_h} we report the PDF $P(H)$ and the cumulative PDF $P(H\geq h)$ of the variable $H$. From both panels, we can note a very similar statistical behaviour and in particular, a clear departure from a local Poisson hypothesis for the events obtained from both dataset. Indeed, the Kolmogorov-Smirnov test gives a probability of $1.7\%$ for Poissonian distributions of WT of both dataset. The average value $\langle H \rangle=0.47 \pm 0.3$ and the shape of $P(H)$ in Figure \ref{p_h}(a) indicate that voids and clusters are both present in a same fashion in both signals.\\
Several observational evidences of clustering in the market volatilities can be found in literature and numerical models \cite{CZ,Pasquini,Gop}. They refer to the clustering volatility phenomenon as a clustering of time periods when volatility fluctuates very strongly and its autocorrelation functions show very long persistent time lags. Several approaches have been guessed to describe this phenomenon, as in Ref.~\cite{KZ}.
Here, by using the test of local Poisson hypothesis for the first time, we can provide a quantitative measure of the phenomenon of volatility clustering in financial time series and we show that voids are also present.
In summary, we have shown that a minimal spin model leads to a fat-tailed distribution function for returns, short-time dependence in the autocorrelation function of returns and, on the contrary, long-term dependence in the volatility signal which shows the typical clustering and ``declustering'' of the real market data.
\acknowledgements{We acknowledge Pierluigi Veltri, Fabio Lepreti and Antonio Vecchio for their helpful suggestions and support. We also acknowledge Sergio Servidio for his continuous interest in a particular aspect of the work. EURO/USD dataset were provided by Gabriella Della Penna.}

\end{document}